\documentclass[twocolumn,floatfix,superscriptaddress,showpacs,preprintnumbers,prl]{revtex4}

\usepackage{graphicx}
\usepackage{epsfig}
\usepackage{amsmath}

\newcommand{\bold}{\bf}

\newcommand{\INOA}{Istituto Nazionale di Ottica Applicata,
Largo E.~Fermi, 6 - 50125 Florence, Italy}

\newcommand{\LENA}{Laboratoire de Neurosciences Cognitives et Imagerie
  C\'{e}r\'{e}brale (LENA) CNRS UPR-640, H\^{o}pital de la
  Salp\^{e}tri\`{e}re. 47~Bd. de l'H\^{o}pital, 75651 Paris CEDEX 13,
  France}

\newcommand{\ITP}{Institut f{\"u}r Theoretische Physik,
  Technische Universit{\"a}t Berlin,
  Hardenbergstra{\ss}e 36, 10623 Berlin, Germany}

\begin{document}

\title{Synchronization in complex networks with age ordering}

\author{D.-U. Hwang}
\affiliation{\INOA}

\author{M. Chavez}
\affiliation{\INOA}
\affiliation{\LENA}

\author{A. Amann}
\affiliation{\INOA}
\affiliation{\ITP}

\author{S. Boccaletti}
\affiliation{\INOA}

\date{\today}

\begin{abstract}

  The propensity for synchronization is studied in a complex network of
  asymmetrically coupled units, where the asymmetry in a given link is
  determined by the relative age of the involved nodes. In growing
  scale-free networks synchronization is enhanced
  when couplings from older to younger nodes are dominant. 
  We describe the requirements for such an effect in a more
  general context and compare with the situations in non growing
  random networks with and without a degree ordering.

\end{abstract}


\maketitle

From the brain over the Internet to human society, complex networks 
are the prominent candidates to describe sophisticated
collaborative dynamics in many areas~\cite{revmod}. 
Complex networks are collections
of dynamical nodes connected by a wiring of edges exhibiting
complex topological properties. Of particular interest are the so
called small-world (SW) and scale free (SF) wirings. SW are
intermediate wirings between regular lattices (RL) and random
networks (RN) \cite{watts98}. They are characterized by a
path-length $\ell$ between any two nodes much shorter than a RL,
but yet a clustering structure much higher than a RN. The basic
property of SW is the logarithmic scaling of $\ell$ with the
network size $N$ ($\ell \propto \log N$), in contrast to the
linear scaling of RL. A specific example of SW is the SF
configuration, in which the degree $k$ (the number of edges in a
node) follows a power law distribution $p(k) \sim k^{-\gamma}$. A SF
network can be grown by adding successively new nodes to the network and
connecting them with the already existing ones by a preferential
attachment rule~\cite{barabasi99}. SW and SF properties were found
in many networks in nature, such as the World Wide Web, the
internet connection wiring, power grids and transportation
networks, neural networks, metabolic and protein networks.

Recently, the dynamics of complex networks has been extensively
investigated  with regard to collective
(synchronized) behaviors \cite{booksPhaseSynchr}, with special
emphasis on the interplay between complexity in the overall
topology and local dynamical properties of the coupled
units~\cite{complNetSync,barahona02,lai03}. Most of the previous works
assumed the local units to be symmetrically connected with uniform
undirected coupling strengths (unweighted links). This
simplification, however, does not match satisfactorily the
peculiarities of many real networks. In ecological systems, for
instance, the non uniform weight in prey-predator interactions
plays a crucial role in determining the food web
dynamics~\cite{maccan98}. Similarly, the interaction between
individuals in social networks  \cite{social} is never symmetric,
rather it depends upon several social factors, such as age, social
class or influence, personal leadership or charisma.

In this Letter, we analyze networks of asymmetrically coupled
dynamical units. By explicitly relating the asymmetry in the
connections to an age order among different nodes, we will give
evidence that age ordered networks provide a better propensity for
synchronization (PFS). In particular we will show that the three
main ingredient maximizing PFS are {\it i)} heterogeneity in the
network topology allowing for the existence of nodes with very
large degrees (hubs) together with nodes with very small degrees 
(non hubs), {\it ii)} asymmetry in the connections forcing a
preferential coupling direction from hubs to non hubs, and {\it
iii)} a structure of connected hubs in the network.

We here adopt the idea that the direction
of an edge can be determined by an age ordering between the
connected nodes. For growing networks (such as SF) the age order
is naturally related to the appearance order of the node during
the growing process. We consider a network of $N$ linearly coupled
identical systems. The equation of motion reads
\begin{equation}\label{systemCoupledOscillators}
    \begin{array}{l l}
        \dot{\bold x}_i = {\bold F}({\bold x}_i)
        - \sigma \sum_{j=1}^{N} G_{ij} {\bold H}[{\bold x}_j], & i
        = 1, \ldots, N,
    \end{array}
\end{equation}
where $\dot{\bold x} = {\bold F}({\bold x})$ governs the local
dynamics of the vector field ${\bold x}_i$ in each node, ${\bold
  H}[{\bold x}]$ is a linear vectorial function, and $\sigma$ is the
coupling strength. $G_{ij} $ is a zero row-sum coupling matrix
with off diagonal entries $G_{ij}=- {\cal A}_{ij}
\frac{\Theta_{ij}}{\sum_{j \in K_i}\Theta_{ij}}$, where ${\cal A}$
is the adjacency matrix, and $\Theta_{ij}=\frac{1-\theta}{2}$
$\left( \Theta_{ij}=\frac{1+\theta}{2} \right)$ for $i>j$ ($i<j$).
$K_i$ is the set of $k_i$ neighbors of the $i^{\text{th}}$ node,
and the parameter $-1 < \theta < 1$ governs the coupling asymmetry
in the network. Precisely, $\theta=0$ yields a symmetric coupling,
while the limit $\theta \to -1$ ($\theta \to +1$) gives a
unidirectional coupling where the old (young) nodes drive the
young (old) ones.
Asymmetric coupling was recently also established for non identical
space extended fields \cite{prljean},
where asymmetry consisted in forcing preferentially the dynamical
regime of a field into the other. 

The network PFS can be inspected by linear stability of the
synchronous state (${\bold x}_i = {\bold x}_s, \forall i$). By
diagonalizing the variational equation, one obtains $N$~blocks of
the form $\dot{\zeta}_i = \text{J}{\bold F}({\bold x}_s) \zeta_i +
\sigma \lambda_i {\bold H} [\zeta_i]$, that only differ by the
eigenvalues of the coupling matrix $\lambda_i$ (here $\text{J}$ is
the Jacobian operator). The behavior of the largest Lyapunov
exponent associated with $\nu=\sigma \lambda_i$ (also called
master stability function~\cite{pecora98}) fully accounts for
the linear stability of the synchronization manifold. Namely, the
synchronous state (associated to $\lambda_1=0$), is stable if all
the remaining blocks, associated to $\lambda_i$ ($i \geq 2$), have
negative Lyapunov exponents. 

For a generic $\theta$, our coupling matrix is asymmetric, and
therefore its spectrum is contained in the complex plane
($\lambda_1=0; \ \lambda_l=\lambda^r_l+j \lambda^i_l, \ l=2,
\ldots, N$). Moreover, since all elements of $G$ are real, non
real eigenvalues appear in pairs of complex conjugates.
In the following we will order the eigenvalues of $G$ for increasing 
real parts.
Gerschgorin's circle theorem \cite{gershgorin} asserts that $G$'s
spectrum is fully contained within the union of circles ($C_i$)
having as centers the diagonal elements of $G$ ($d_i$), and as
radii the sums of the absolute values of the other elements in the
corresponding rows ($\{ \lambda_l \} \subset \cup_i C_i \left[
d_i,\sum_{j \neq i} \mid G_{ij} \mid  \right]$).

By construction, the diagonal elements of $G$ are normalized to 1
in all possible cases. It is crucial to emphasize the physical and
mathematical relevance of this choice. Physically, this
normalization prevents the coupling term from being arbitrarily
large (or arbitrarily small) for all possible network topologies
and sizes, thus making it a meaningful realization of what happens
in many real world situations  (such as neuronal networks) where
the local influence of the environment on the dynamics does not
scale with the number of connections. Mathematically, since $G$ is
a zero row-sum matrix (and furthermore $d_i=\sum_{j \neq i} \mid
G_{ij} \mid$ because all non zero off diagonal elements are
negative), this warrants {\it in all cases} that $G$'s
spectrum is fully contained within the unit circle centered at 1
on the real axis ($\mid \lambda_l-1 \mid \leq 1, \ \forall l$),
giving the following inequalities: {\it i)} $0 < \lambda^r_2 \leq
\ldots \leq \lambda^r_N \leq 2$, and {\it ii)} $\mid \lambda^i_l
\mid \leq 1, \ \forall l$. This latter property is essential to
provide a consistent and unique mathematical framework within
which one can formally assess the relative merit of one topology
against another for optimal PFS {\it independent of} 
the network size or the local dynamics.

\begin{figure}[htbp]
  \centering
  \includegraphics[height=5cm]{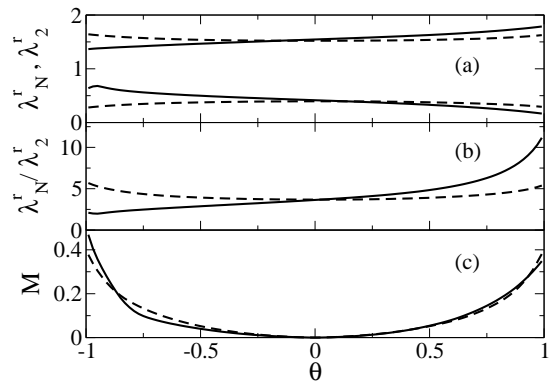}
  \caption{$\lambda^r_N$ and $\lambda^r_2$ (a),
  $\frac{\lambda^r_N}{\lambda^r_2}$(b), and $M$ (c) {\it vs}. $\theta$
  for SF ($m=5$ and $B=0$, solid lines) and RN (dashed lines).
All quantities and all other network parameters are specified in
the text.}
  \label{fig:eigvalues}
\end{figure}

Let $\cal R$ be the bounded region in the complex plane where the
master stability function provides a negative Lyapunov
exponent. The stability condition for the synchronous state is
that the set $\{ \sigma \lambda_l , l=2,\ldots,N \}$ be entirely
contained in $\cal R$ for a given $\sigma$. The best PFS is then
assured when both the ratio $\frac{\lambda^r_N}{\lambda^r_2}$ and
$M \equiv \max_l \{ \mid \lambda^i_l \mid \}$ are simultaneously
made as small as possible.

With the help of such stipulations, we start with analyzing the
effects of heterogeneity in the node degree distribution. This is
done by comparing the PFS of a class of SF networks with different
degree distributions with a highly homogeneous RN. The used class
of SF networks is obtained by a generalization of the preferential
attachment growing procedure~\cite{barabasi99}. Namely, starting
from $m+1$ all to all connected nodes, at each step a new
node is added with $m$~links, connecting to old nodes with
probability $p_i = \frac{k_i+B}{\sum_j(k_j +B)}$ ($k_i$ being the
degree of the $i^{\text{th}}$ node, and $B$ a tunable real
parameter, representing the initial attractiveness of each node).
The $\gamma$ exponent of the power law scaling in the degree
distribution $p(k) \sim k^{-\gamma(B,m)}$ is then given by
$\gamma (B,m) = 3 + \frac{B}{m}$ in the thermodynamic ($N \to
\infty$) limit \cite{netInNature}. While the average degree is by
construction $\langle k \rangle = 2m$ (thus independent of $B$),
the heterogeneity of the degree distribution can be strongly
modified by $B$. This induces convergence of higher order moments
of $p(k)$, in contrast with the case $B=0$ that recovers the
original preferential attachment rule~\cite{barabasi99}.

For comparison, a highly homogeneous Erd\"os-R\'enyi RN
\cite{erdos} is considered, with connection probability
$P=\frac{2m}{N-1}$ (giving same average degree $\langle k \rangle
=2m$), with an arbitrary initial age ordering. Fig.
\ref{fig:eigvalues}a) shows $\lambda^r_N$ and $\lambda^r_2$ {\it
vs.} $\theta$ for SF with $m=5$ and $B=0$ (solid line) and for the
chosen RN (dashed line). All calculations refer to ensemble
averages over 24 different realizations of networks with 500
nodes. For RN, the curve $\lambda^r_N (\theta)$ [$\lambda^r_2
(\theta)$] displays a minimum [a maximum] for $\theta=0$, showing
that asymmetry here deteriorates the network PFS. At variance, for
SF the difference between $\lambda^r_N$ and $\lambda^r_2$
continuously shrinks, as $\theta$ decreases. As a consequence,
Fig. \ref{fig:eigvalues}b) reports the behavior of the eigenratio
$\frac{\lambda^r_N}{\lambda^r_2}$, making it clear that while the
best PFS in RN is obtained for the symmetric case ($\theta=0$), SF
shows better (worse) PFS in the asymmetric case for $\theta \to
-1$ ($\theta \to 1$). As for the imaginary part of the spectra,
Fig. \ref{fig:eigvalues}c) reports $M$ {\it vs.} $\theta$,
indicating only very small differences between SF and RN in the
whole range of the asymmetry parameter, and highlighting that the
contribution to PFS of the imaginary part of the spectra does not
depend significantly on the specific network structure. These
findings have been consistently observed in subsequent results.

\begin{figure}[tbp]
  \centering
  \includegraphics[height=6cm]{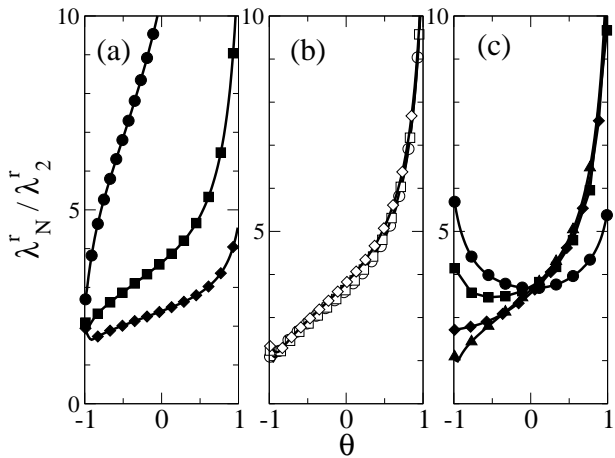}
  \caption{$\frac{\lambda^r_N}{\lambda^r_2}$ {\it vs.}
  $\theta$ for (a) SF with $B=0$ and $m=2$ (circles), 5 (squares), 10 (diamonds); (b)
SF with $m=5$ and $B=0$ (circles), 5 (squares), and 10 (diamonds);
(c) RN with arbitrary age order (circles), RN with age depending
on degree (squares),  SF with $m=5$ and $B=0$ without connection
between nodes 1 and 5 (diamonds), and SF with $m=5$ and $B=0$
(triangles).}
  \label{fig:eigenratio}
\end{figure}

The second step of our study is the investigation of PFS in SF at
different $m$ and $B$ values. In Fig. \ref{fig:eigenratio} (a),
PFS is compared for different $m$ for $B=0$. Since $m$ determines
the average degree, as $m$ increases, the average connectivity
increases and synchronizability is naturally enhanced. The most
important point is that the monotonically decreasing behavior of
$\frac{\lambda^r_N}{\lambda^r_2}$ with $\theta$ persists for all
$m$, indicating that synchronization is always enhanced when
$\theta$ becomes smaller. In Fig. \ref{fig:eigenratio} (b), PFS is
compared for $m=5$ and for various values of $B$, altering
the exponent of the degree distribution. For all $B$ values, the
same enhancement for negative $\theta$ was observed. Therefore, one can
conclude that in aged growing networks PFS depends on the average
degree, but the asymmetry enhanced synchronization phenomenon does
not depend on heterogeneity.

\begin{figure}[tbp]
  \centering
  \includegraphics[height=9cm]{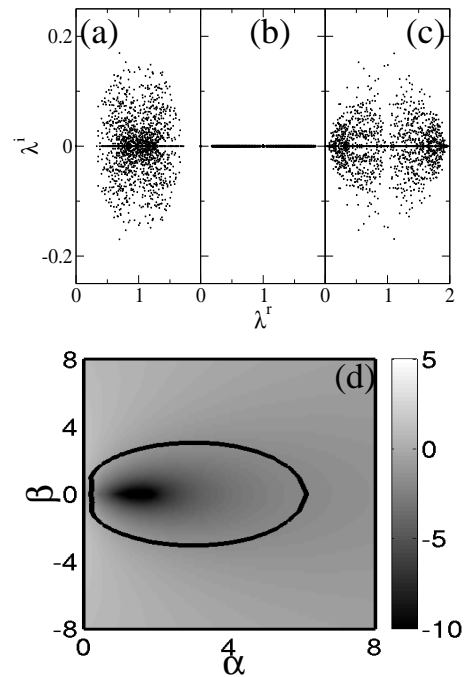}
  \caption{Distribution of $G$'s eigenvalues in the complex
  plane for SF with $m=2$, $B=0$ and
   $\theta=-0.8$ (a), $\theta=0$ (b), and $\theta= 0.8$ (c).
 (d) Master stability function in the complex plane for coupled
R\"ossler oscillators. } \label{fig:master}
\end{figure}

This leads us to discuss the main point of our study, concerning
the determination of the essential topological ingredients
enhancing PFS in weighted (aged) networks. The first ingredient is
that the weighting must induce \emph{a dominant interaction
from hub to non hub nodes}. This can be easily understood by a
simple example: the case of a star network consisting of a single
large hub (the center of the star) and several non-hub nodes
connected to the hub. When the dominant coupling direction is from
the non-hub nodes to the hub node, synchronization is impossible
because the hub receives a set of independent inputs from the
different non-hub nodes. In the reverse case (when the center
drives the periphery of the star) synchronization can be easily
achieved. The very same mechanism occurs in our SF case. Indeed,
for positive (negative) $\theta$ values, the dominant coupling
direction is from younger (older) to older (younger) nodes. Now,
in SF the minimal degree of a node is by construction $m$ and
older nodes are more likely to display larger degrees than younger
ones, so that a negative $\theta$ here induces a dominant coupling
direction from hubs to non-hub nodes.

The second ingredient is that \emph{the network contains a
structure of connected hubs influencing the other nodes}. In our
SF case, the normalization in the off diagonal elements of $G$
\cite{norma} assures that hubs receive an input from a connected
node scaling with the inverse of their degree, and therefore the
structure of hubs is connected always with the rest of the network
in a way that is independent on the network size.

In order to make evident the validity of these claims, we have
performed a careful analysis on enhanced PFS over a series of
ad-hoc modified networks. The results are summarized in Fig.
\ref{fig:eigenratio} (c). First, we have reordered the node age in
RN according to each node degree. The resulting
$\frac{\lambda^r_N}{\lambda^r_2} (\theta)$  (curve with squares)
shows now a minimum for an asymmetric configuration
($\theta\approx-0.5$), in contrast to the case with arbitrary
aging (curve with circles). This confirms the need of a dominant
interaction from hubs to non hubs for improving PFS, also for
highly homogeneous networks. As for the second ingredient,
starting from a SF with $m=5$ and $B=0$ (curve with triangles), we
artificially disconnected the initially existing link between the first and
fifth network nodes. These are indeed the two hubs with highest
degree in the SF configuration. The result is shown in the curve
with diamonds, where one sees that such a small perturbation (the
difference in the two networks is limited to only a link) is
already sufficient to substantially weaken the PFS. The situation
remains however better than the two RN cases, indicating that the
structure of growing aged network inherently enhances
synchronization.

\begin{figure}[tbp]
  \centering
  \includegraphics[height=6cm]{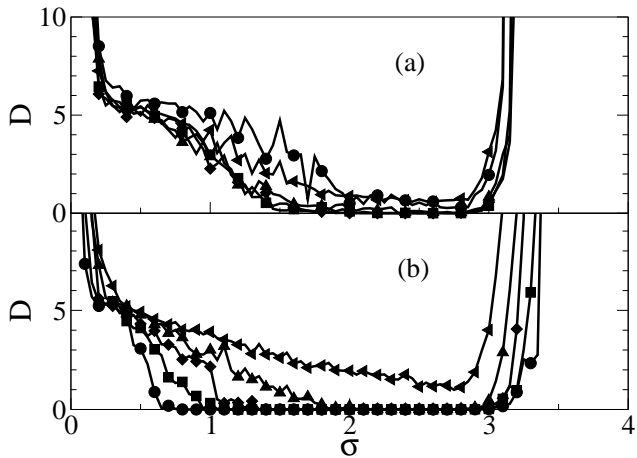}
  \caption{$\langle D \rangle$ (see text for definition) {\it vs.} $\sigma$ for RN (a) and
  for SF ($m=2,B=0$, b). In both graphs the curves with circles, squares, diamonds, up
  triangles, and left triangles refer to $\theta=~-0.8,~-0.4,~0,~0.4,~0.8$,
  respectively.}
  \label{fig:rosslerdiff}
\end{figure}

Finally, we prove the validity of our arguments by applying them
to networks of coupled chaotic R\"ossler oscillators \cite{roe}.
The dynamics is ruled by Eq. \ref{systemCoupledOscillators}, with
${\bold x} = (x,y,z)$, ${\bold F}({\bold x})=[-y - z, \ x + 0.165
y, \ 0.2 + z(x-10)]$, and ${\bold H}[{\bold x}]=x$. The master
stability function is depicted in Fig. \ref{fig:master} (d) in the
complex plane $\nu=\alpha+j\beta$. The bold solid line (denoting a
zero Lyapunov exponent) is the boundary between stability
($\cal R$) and instability regions for the synchronization
manifold. The whole network synchronizes when all the eigenvalues
of $G$ (multiplied by $\sigma$) locate inside $\cal R$. Figs.
\ref{fig:master} (a,b,c) report the location of the $G$ spectrum
for SF for $m=2, B=0, N=500$ and with $\theta=-0.8, \ 0,$ and
$0.8$, respectively. While for $\theta=0$ the spectrum is real,
comparison of (a) and (c) shows that the spectra at negative
$\theta$ values are much less dispersed in the complex plane, thus
increasing the range of $\sigma$ values for which synchronization
can be achieved in the network.

The appearance of the synchronous state can be monitored by
looking at the vanishing of the time average (over a window $T$)
synchronization error $ \langle D\rangle =
\frac{1}{T(N-1)}\sum_{j>1}
  \int_t^{t+T} ||\mathbf{x}_j-\mathbf{x}_1||dt'$. In the present case, we
  adopt as vector norm $||\mathbf{x}||=|x|+|y|+|z|$.
Fig. \ref{fig:rosslerdiff} reports $\langle D\rangle $ {\it vs}.
$\theta$ for RN with arbitrary age (a) and for SF with $m=2, B=0$
(b). The curves with circles, squares, diamonds, up
triangles, and left triangles refer to
$\theta=~-0.8,~-0.4,~0,~0.4,~0.8$, respectively. While in the RN
case the range for synchronization is substantially independent on
$\theta$ [reflecting the behavior of
$\frac{\lambda^r_N}{\lambda^r_2}$ in the dashed line of Fig.
\ref{fig:eigvalues} (b)], the case with SF [to be compared with
the curve with circles in Fig. \ref{fig:eigenratio} (a)] confirms
that synchronization is strongly affected by the asymmetry. In
particular, negative (positive) values of $\theta$ have the
effects of increasing (decreasing) the range of coupling strengths
over which synchronization occurs with respect to the symmetric
case $\theta=0$.

In conclusion we have demonstrated that PFS is enhanced in aged
networks of asymmetrically coupled units. In growing SF such
enhancement is particularly evident when the dominant coupling
direction is from older to younger nodes. Our study allows to
individuate the main network topological ingredients at the basis
of synchronization enhancement. A key aspect of social
organizations is the dynamics of information exchange. Our
approach may provide new insights in the study of collective
communication or coordination in distributed social networks, as
well as useful hints for understanding the formation of social
collective behaviors (leading opinions, rumors, political
orientations, dominant tastes, habits, fashion). Work partly
supported by MIUR-FIRB project n. RBNE01CW3M-001.


\begin{thebibliography}{99}
\bibitem{norma}
For SF growing networks, the non zero off diagonal elements of the
coupling matrix are given by $G_{ij}= \frac{1 \mp \theta}{k_i -
\theta [k_i - 2 \min{(i-1,m)}]} $, where $\mp$ stands for $i>j$ and
$i<j$, respectively.
\end{thebibliography}
\end{document}